# Replication of filtered interferometer measurements in interstellar communications

*William J. Crilly Jr.*

Green Bank Observatory, West Virginia, USA

*Abstract*— Interstellar communication signals have been conjectured to be present, albeit difficult to identify. Experiments conducted since 2018 indicate an anomalous presence of a type of speculated interstellar signal, Δt Δf polarized pulse pairs, thought to be possibly sourced from a celestial direction near 5.25 hr Right Ascension and -7.6º Declination. A recent experiment utilizing a radio interferometer identified anomalous pulse pairs associated with these celestial coordinates. The experiment is described in arXiv:2404.08994. Other experiments produced anomalous results, reported in arXiv:2105.03727, arXiv:2106.10168, arXiv:2202.12791 and arXiv:2203.10065. After the recent experiment was concluded, the interferometer antenna elements were modified to have increased aperture and reduction in radio interference-caused false positives. An experiment was conducted to attempt replication of the previously reported interferometer measurements. Observations are reported here. Apparent replicated falsification of an expected random white noise explanatory hypothesis compels the development and testing of alternate and auxiliary hypotheses.

## I. INTRODUCTION

An important step in hypothesis falsification is the replication of measurement results.[1] Time-varying effects generally confound one's ability to develop explanations for unusual phenomena. Given that previous experimental reports have produced unexpected results, [2][3][4][5][6], a long term examination of the interferometer experiment described recently in [2] is underway. The hypothesis tested in this work is based on an Additive White Gaussian Noise (AWGN) explanatory model. In Bayesian inference, a replication measurement value is defined to help one understand explanatory variables.[7] The presence, or absence, of replicated measurement anomalies provides a process to refine prior models that explain phenomena.

The Search for Extraterrestrial Intelligence (SETI) using radio is a complex endeavor. Fortunately there has been extensive study of energy-efficient and discoverable communication systems, and the development of models that predict the likelihood of narrow bandwidth signals in random noise. This work helps experimenters reduce the signal search space.[8][9][10][11][12][13]

## II. HYPOTHESIS

The hypothesis tested in this experiment is similar to the hypothesis tested in [2]. Changes to the hypothesis include the addition of replication and pulse pair candidate filter qualifiers.

*Hypothesis:* Narrow-bandwidth energy-efficient interstellar communication signals are expected to be explained, and replicated, during multiple observation runs, by an AWGN signal reception model, while using an observer meridian pointing interferometer, and receiver design that reduces RFI and provides Signal-to-Noise Ratio (SNR) threshold reduction. Receiver processing and filters provide for the selection of Δt Δf polarized pulse pair candidates and measurements associated with low values of differential angles-of-arrival. Algorithms measure Δt Δf polarized pulse pairs having a Right Ascension (RA) range centered on a prior anomalous celestial direction: 5.25 ± 0.15 hr RA and -7.6° ±1° declination (DEC). In the current work, the time difference Δt, between narrow bandwidth pulses in a pair, is nominally equal to zero seconds.

## III. METHOD OF MEASUREMENT

*Objectives*

The search for interstellar communication signals naturally leads to the need for extensive development and testing of a large number of possible explanatory hypotheses. Replication of unusual anomalous measurement results provides a mechanism to develop explanatory hypotheses, together with providing a means of verifying measurement instruments. A goal of this work is to design and implement measurements and filters that seek to find and store potential communication signals that may be described, filtered, replicated and temporally studied.

*Observation runs*

In this report, hypothesis testing is performed using data from two observation runs, labelled **O5** and **O6**. Observation runs are numerically indexed by the five associated chronological arXiv announced reports relevant to this project. The observation run **O5** processes pulse pair data [2] collected during 61 contiguous days. A new observation run presented in this report, **O6**, processes raw pulse pair data collected during 46 contiguous days.

*Antenna modifications*

The interferometer elements used in the experiment described in [2] used rectangular aperture extensions to reduce the system noise contribution due to feed antenna spillover. The extensions on each element were removed and replaced with circular paraboloid surface extensions, resulting in a circular offset-fed aperture diameter of ten feet. The feeds were not modified or adjusted. Continuum signal power to noise power ratio was measured at 0.18 dB in **O5** and 0.25 dB in **O6**, using astronomical object NRAO 5690 [14] as a source, while applying a Gaussian signal plus flat noise modeled antenna element and source response.

---
William J. (Skip) Crilly Jr. is a Volunteer Science Ambassador in Education & Public Outreach of the Green Bank Observatory. email: wcrilly@nrao.edu

*Antenna calibration*

Following antenna improvements, astronomical object NRAO 5690 was measured to determine element antenna beam center pointing and beam width calibration.

Measurements of received power of NRAO 5690, indicated a 0.065 hr RA increase in **O6** peak continuum, compared to **O5** archived continuum data. Full Width Half Maximum (FWHM) interferometer element antenna responses were calculated using Gaussian plus flat noise estimates. The peak NRAO 5690 continuum response, together with the Gaussian plus flat noise estimates, were used to estimate composite FWHM antenna power response during **O5** at 9.0º and **O6** at 8.2º RA angle, at -8º DEC. The estimated antenna FWHM is displayed on the horizontal axes in **Figs. 1** and **2.**

The center RA of FWHM displayed in **Figs. 1** and **2** was set to 5.25 hr, based on previous 180º azimuth RA calibration in **O5** work, while the **O6** azimuth in post-acquisition processing was set to 181.5º, after antenna modification, NRAO 5690 measurement and adjustment for observation epoch.

*Cross-correlator instrument delay*

Cable delays were added in the East element RF path, to test the broadband complex correlator, after antenna element surface modification, and to adjust peak response to delay bins that avoid common mode correlator response. The broadband complex correlator is not used in the reported measurements of this experiment. Rather, 3.7 Hz bandwidth pulse RF phase difference measurements are performed, calculated from the complex amplitude of an FFT bin voltage. The RF phase difference contributes measurements to estimate differential angle-of-arrival values of measured pulse pairs. Broadband correlator measurements are expected to be useful, due to low natural astronomical object sensitivity using the pulse pair detection in this experiment, partially resulting from the $1/\sqrt{\Delta v\, t\, n}$ radio telescope sensitivity factor **[15]** equal to 1, with $\Delta v$=3.7 Hz, $t$= 0.27 s, and $n$=1.

The instrument delay was measured by seeking the correlator delay tap having the most sensitive measurement of expected correlator phase of NRAO 5690. Examination of **O5** NRAO 5690 observations yielded the **O5** $\tau_{INT}$ at -16±10 ns and the **O6** $\tau_{INT}$ at -144±10 ns. The $\tau_{INT}$ values of each observation were adjusted within these ranges to maximize the Cohen's **d** values within the FWHM, shown in **Figs. 1** and **2.**

*Measurement filters*

The measurements reported here use filter settings as reported in **[2]**, with exceptions:

1. The **Log$_{10}$ Δf / MHz** filter upper limit value was increased from the value of -1.0 in the **O5** report **[2]** to +0.3 in the current report. **Δf** filtered values have a range of 7.9 Hz to 2.0 MHz. The logarithm is included to ease event likelihood calculation, **[3]** Appendix C, in future planned work.

2. The RF frequency upper limit was increased from 1435 MHz to 1455 MHz, while the lower limit at 1405 MHz and the zero-IF RFI rejection filter at 1424 to 1426 MHz were unchanged.

3. RA bin indexes were used in figures in **[2]**. In the current presentation, the horizontal axis of the presented figures is the peak response pointing RA, corresponding to azimuth-corrected local meridian transit times, calculated from a GPS clock measurement captured synchronously with analog telescope signals. RA pointing values are thought to yield more presented information than RA bin index values.

4. The number of candidate pulse pairs, used for statistical power calculations, increased approximately a factor of ten, due to the increased **Log$_{10}$ Δf /MHz** range and increased RF frequency range.

5. The $\Delta_{\Delta f}\Delta_{EW}\Phi_{INT.RF}$ filter was adjusted from a limit of ± 0.1 radian in **[2]** to ± 0.04 radian, to seek pulse pairs concentrated near zero phase, expected to be associated with low corrected differential angles-of-arrival.

Measurement filter values are described in Table 1.

| Figure | Observation | $\Delta_{\Delta f}\Delta_{EW}\Phi_{INT.RF}$ (radians) | Log$_{10}$ Δf /MHz |
|---|---|---|---|
| 1 | **O5** | 0.0 ± 0.04 | -5.1 to +0.3 |
| 2 | **O6** | 0.0 ± 0.04 | -5.1 to +0.3 |

**Table 1:** The same second level filters process data from two observing runs.

*Changed definition of the ΔΔ operator in* $\Delta\Delta\Phi_{INT.RF}$

The measurement $\Delta\Delta\Phi_{INT.RF}$, **[2]** has **ΔΔ** explicitly subscripted in this work, and is defined, with an added Δf related factor, to that in **[2]**, as follows,

$$\Delta_{\Delta f}\Delta_{EW}\Phi_{INT.RF}(i) = (\Phi_{INT.RF.WEST}(i) - \Phi_{INT.RF.EAST}(i))$$
$$-(\Phi_{INT.RF.WEST}(i-1) - \Phi_{INT.RF.EAST}(i-1)) + 2\pi\, \Delta f\, \tau_{INT}, \quad (1)$$

where $\Phi_{INT.RF}(i)$ is the interferometer RF phase measurement, per FFT bin, per element, in radians, Δf is pulse pair component frequency spacing in Hz, and $\tau_{INT}$ is instrument delay in seconds. The count index *i* increments with the FFT bin-sorted, *i*-indexed filtered polarized pulse pair candidates, having SNR East > 8.5 dB and SNR West > 8.5 dB.

*Δt values in candidate pulse pairs*

In the previous interferometer report **[2],** the description of the definition of $\Delta\Delta\Phi_{INT.RF}$ states the following: "The count index *i* increments with the FFT bin-sorted SNR East > 8.5 dB and SNR West > 8.5 dB Δt=0 polarized pulse pair candidates."

The statement above regarding Δt=0 is accurate only if the FFT input data acquisition time did not change between the Δf component pulses that are within filter limits. The possibility that Δt > 0 exceptions are present in second-level filtered measurements was tested by examining samples of Δt values after second-level filtering. In a test of 2,577 sorted candidate samples, one pulse pair measured Δt=2.75 s and four pulse pairs measured Δt=0.25 s. Candidates having Δt > 0 are not excluded in the processing in this work.

IV. OBSERVATIONS

*Measurement of effect size across RA populations*

**Figs. 1** and **2** plot **O5** and **O6** Cohen's **d** = ΔMean / std.dev.**[16]** of pulse pair counts, using binomial model likelihood **[7]**, in 0.1 hr ΔRA bins vs RA (hrs). ΔMean is the measured count value minus the expected count value, the latter calculated from binomial model statistics. ΔMean is then normalized by the binomial model standard deviation, calculated given the trial count and event probability.

Replication of filtered interferometer measurements in interstellar communications

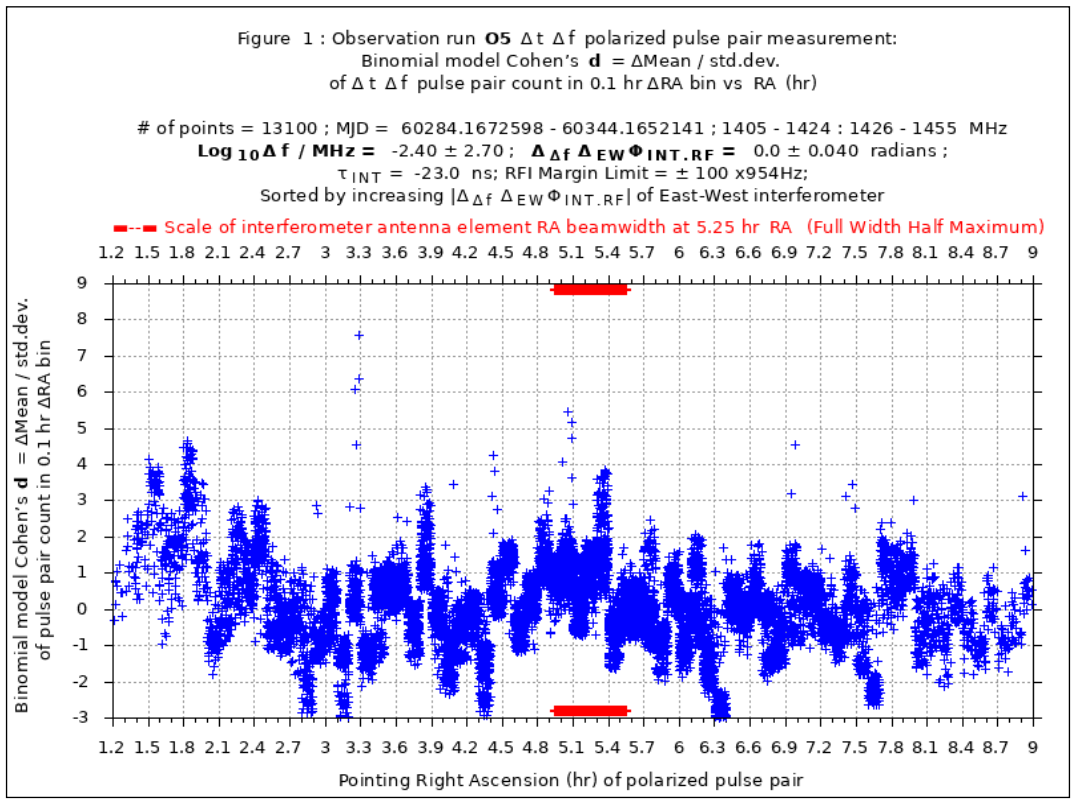

**Figure 1:** Anomalous Δt Δf pulse pair counts are observed in the FWHM of the interferometer instrument, at higher levels of Cohen's **d** in the current work, than observed in **[2]** Fig. 2. Binomial cumulative probability: e.g. (328 trials, 8.2 mean, count > mean, at 3.8 s.d.) = 0.0003.

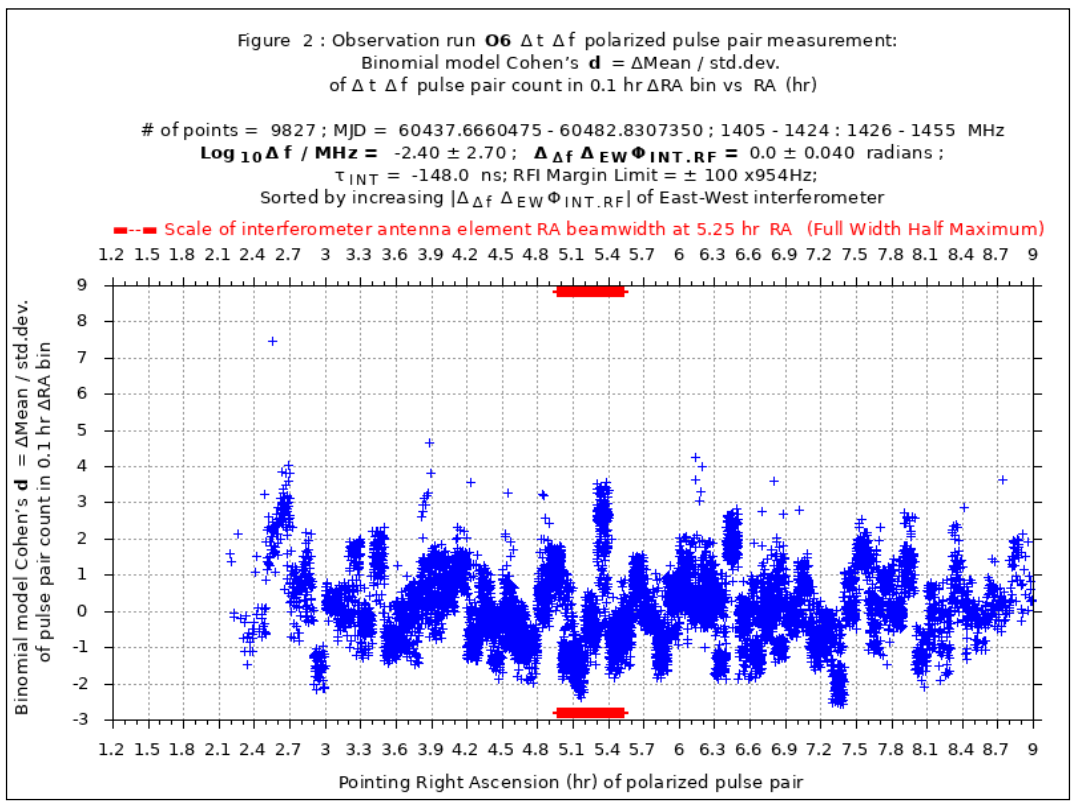

**Figure 2: O6** measurements indicate anomalies in the Δt Δf pulse pair counts, within the FWHM beam width. Binomial cumulative probability: e.g. (246 trials, 6.1 mean, count > mean, at 3.5 s.d.) = 0.0009.

## V. Discussion

The 5.25 ± 0.15 hr RA, -7.6° ± 1° *DEC* celestial direction has been associated with a significant number of anomalous pulse pair events in experimental work, since 2018.**[2][3][4][5][6]** In the current work, the interferometer instrument filters are adjusted to maximize the detection of near zero valued pulse pair Δf component difference in angles-of-arrival, derived from delay corrected narrow bandwidth RF phase measurements across interferometer elements.

Presumably, intentional communication signal simultaneous components may be associated with each other if they measure a low value of difference in angle-of-arrival. The use of an adjustable filter to seek these measurements might be considered to be an ad-hoc choice, albeit only in the width of the filter, since the filter center is set to zero radians, a choice set by theory.

### Using RF phase to seek angle-of-arrival

Difficulties arise when using one measurement to derive an estimate of another quantity. For example, a high RF frequency difference, Δf, of the pulse pair components, induces a residual RF phase difference that requires careful metrology. Another issue is the aliasing of arrival direction measurements, due to n·2π RF phase wrapping.

On the other hand, narrow bandwidth pulses, transmitted with various bandwidths and frequency spacings, facilitate a reduction in angle-of-arrival measurement uncertainty induced by n·2π RF phase wrapping. These types of signals are expected to be transmitted in a high capacity energy efficient communication system intended to be discoverable without undue effort.

### Increase in sample population sizes

The observations in this report use filter adjustments that greatly increase the sample population size, compared to the population size in **[2]**. Archived **O5** first level filtered data was applied to these wider filters, and the same filters were used to process **O6** first level filtered data. Generally, wide filters and large population sizes are favored when attempting to apply statistical analysis to model falsification.

In text below **Figs. 1** and **2**, large sample size calculated estimates of the binomial distribution likelihood of high values of Cohen's **d** are presented. The example sample size in these calculations is set by the product of the total number of points in the RA bin data and the binomial event probability in the RA bin, nominally 0.1 hr / 4.0 hr = 0.025. Precise calculation of RA bin probability for each bin is computed from first level filtered data applied to the second level signal processing.

Each observation run, **O5** and **O6**, calculates to have a binomial model likelihood below 0.001. The possibility that a 61 day observation run and a 46 day observation run would both have these low likelihoods, due to random chance and noise, seems remote.

### Second level filter adjustments

In this work, the $Log_{10}$ Δf /MHz < -1.0 filter setting used in **[2]** did not show a high response in Cohen's **d** in the FWHM, when **O6** first level filtered data was applied to this prior filter. This was a peculiar finding and implies an absence of replication at the prior filter setting. Reasoning for this leads to speculation about hypothetical transmitter characteristics. For example, low values of Δf do not need to be continuously transmitted from an intentionally discoverable transmitter.

In this work, the second level filters were adjusted to increase the Cohen's **d** values in the FWHM, of **O5** and **O6**. Results of these filter adjustments are shown in **Figs. 1** and **2**.

The $\Delta_{\Delta f}\Delta_{EW}\Phi_{INT.RF}$ filter was adjusted, in its range around zero radians, to look for peak response in Cohen's **d**. There is a possibility that chance results are being observed, due to the intentional peaking of response with filter adjustments.

Further replication attempts, data examination, model development and statistical analysis ameliorate the issue of filter adjustment causing random chance results.

### RA bin value similarity between O5, O6 and O1

**Figs. 1** and **Fig. 2** indicate a highly unlikely concentration of pulse pairs within the FWHM, during **O5** and **O6**, in the same RA bin, near the estimated FWHM centers. In both **O5** and **O6** measurements, the Cohen's **d** measured peak sustained values at the highest levels observed within the range of 2.7 to 9 hours RA, during 61 day and 46 day observations.

A continuously transmitting celestial source would be expected to produce receiver response near the most sensitive antenna pointing, i.e. the central region of each interferometer element antenna. It seems that other pointing directions within the FWHM should show similarly high Cohen's **d** values. These RA adjacent values appear in **Fig. 1** while appearing absent in **Fig. 2**. The Cohen's **d** measurement of pulse pair count appeared to be sensitive to the adjusted settings of instrument delay $\tau_{INT}$ and $Log_{10}$ Δf /MHz, possibly explaining low adjacent values within the FWHM.

In the **O5** previous experimental report in **[2]**, Fig. 2, Cohen's **d** high values appeared in the 5.2 to 5.3 hr RA bin. Several high values of Cohen's **d**, in the 5.3 to 5.4 hour RA bin, were observed in **[2]** Fig. 2. Two points in the 5.3 to 5.4 hour bin had values greater than Cohen's **d** = 4. In the current report, **O5** data shows peak value in the 5.3 to 5.4 hour RA bin. The reason for a discrepancy is not understood. There is a possibility that inclusion of $\tau_{INT}$ in the calculation of $\Delta_{\Delta f}\Delta_{EW}\Phi_{INT.RF}$, together with the use of a wider setting of $Log_{10}$ Δf /MHz in this work, affected the **O5** results. Further work is required to understand this.

In the **O1** report **[3]** Fig. 22, Δt=0 , |Δf| < 400 Hz pulse pairs were observed, in a 164 beam transit experiment using a 26 foot diameter radio telescope, at lowest binomial model density, in the RA range from 5.17 to 5.39 hours. This prior determined RA range overlaps most of the 5.3 to 5.4 hour RA range, anomalous in the recent **O5** and **O6** observations.

### Detectability given a bursty transmitting source

Interferometer response may be expected in the presence of intentionally transmitted aperiodic signals, given communication theory concepts.

1. Transmitter detectability is enhanced if signals are transmitted within a short duration. Storing energy, then releasing the energy, in a numerous set of narrow bandwidth pulses, reduces the AWGN model calculated likelihood of such composite events.

2. The decoding to information bits of discovery signals is enhanced, as the pulses may be distinguished from noise by their anomalous presence within a short time duration.

3. Available transmitter energy may be used to transmit at separate times in separate directions.

The temporal nature of the pulse pair count anomalies has not been studied in this work.

*Ideas that might explain anomalous pulse pair counts*

1. A metrology issue may be present in the estimated beam pointing, possibly changing over time.

2. Transmission of concentrated bunches of pulses might be aliasing, or may be synchronized with the rotation and/or revolution of the Earth, over multiple days, perhaps intentionally.

3. Receiver filters, intentionally adjusted for maximum response, may show responses caused by random noise.

4. Natural astronomical objects are often bursty, albeit not readily observable in separate interferometer element SNR measurements in narrow (< 10 Hz) signal bandwidths.

5. The observation of pulse pair anomalies in multiple RA directions has not been studied in this work, and may lead to alternate and auxiliary hypotheses.

6. RFI may have not been excised by the RFI identification and excision algorithm.

7. Software and/or algorithm errors may be present.

8. One or more unknown transmitters may be present.

*SNR measurement*

The measurement of SNR in this work is calculated from the ratio of measured power in a 3.7 Hz bandwidth to the expected average power in a 3.7 Hz bandwidth, the latter derived from a 954 Hz bandwidth measurement, i.e. 256 FFT bins surrounding the 3.7 Hz bandwidth. Signal and noise power measurements are averaged during the same 0.27s time interval of the FFT input data. The SNR measurement method used in this work balances the need to provide an estimate of the noise power that has low residual fluctuations, with the need to reject time and RF frequency dependent RFI that affects the noise and signal measurements in generally unpredictable ways.

In the SNR measurement, the expected distribution of high SNR measurements may be calculated using a Rayleigh amplitude distribution **[2][3][5][15]**, assuming flat spectral noise power density, and no narrowband components are present. This distribution is expected to be independent of AWGN noise power, providing there are no confounding issues. In general, the SNR measurement provides a natural mechanism to excise emitters that are modeled as flat bandwidth emitters across a 954 Hz RF bandwidth.

Difficulties arise when trying to determine the cause of an SNR threshold crossing event. Examples follow.

1. The 954 Hz bandwidth measurement may be briefly low, while the 3.7 Hz noise may be a Rayleigh distributed high amplitude noise-caused outlier. This scenario is unlike that expected from a bursty natural source.

2. The 3.7 Hz FFT bin may be near the edge of the 256 bin segment used to calculate and estimate noise power. The effect of adjacent segments is not examined in this work.

3. The celestial sky has a variety of natural short duration fluctuations that have the potential to confound SNR measurements.

4. The expected Doppler spread of rotating natural astronomical objects generally implies an absence of 3.7 Hz bandwidth signal components from narrow bandwidth natural processes, e.g. MASERs. This conjecture needs to be tested using astronomical object and measuring instrument models.

*Phase measurement*

The effect of system noise on the uncertainty of the measurement of interferometer differential RF signal phase depends on the signal model used, e.g. Ricean, or Rayleigh distributed amplitude statistics. Work is underway to produce theory and simulations, to model and test the expected uncertainty of phase measurements in the experimental interferometer system. This work may lead to a theory-based setting of the $\Delta_{\Delta f}\Delta_{EW}\Phi_{INT.RF}$ filter, ameliorating the issue of adjustment-induced measurement response.

*Absence of independent corroboration*

The author is not aware of experiments performed similar to those described here. Absence of reported independent corroboration leads to questions about the confidence of conclusions.

## VI. CONCLUSIONS

The AWGN model in this hypothesis continues to not provide an explanation of the measurement of narrow bandwidth $\Delta t$ $\Delta f$ polarized pulse pairs, observed in the direction of 5.25 hr RA, -8° DEC. **Figs. 1** and **2** present measurements that replicate the presence of anomalies within the FWHM of the interferometer elements, during two long duration observation runs. Replication of falsifying experimental results reduces support for a hypothesis. Further examination of pulse pairs using interferometers and other methods are needed to better understand the phenomena.

## VII. FURTHER WORK

1. Conduct ongoing interferometer experiments, seeking characteristics of replication and the sensitivity of measurement filters.

2. Study natural object and intentional transmitter models, including the measurement effect of uncorrelated and correlated noise across interferometer elements.

3. Study the short and long term temporal characteristics of measurements.

4. The inclusion of $\Delta t > 0$ in the measurement filters seems important, as it appears to be useful in the development of source models.

5. Study the measurement characteristics of the differential RF phase filter.

6. Add a third interferometer element.

7. Prioritize tasks with previously proposed further work.

## VIII. ACKNOWLEDGEMENTS